\documentclass[aip,reprint]{revtex4-1}
\usepackage{graphicx}

\begin{document}


\title{Study of accelerated ion energy and spatial distribution with variable thickness liquid crystal targets} 



\author{P. L. Poole}
\email[]{poole.134@osu.edu}
\author{C. Willis}
\author{C. D. Andereck}
\author{L. Van Woerkom}
\author{D. W. Schumacher}
\affiliation{The Ohio State University}


\date{\today}

\begin{abstract}
We report on laser-based ion acceleration using freely suspended liquid crystal film targets, formed with thicknesses varying from 100 $nm$ to 2 $\mu m$ for this experiment. Optimization of Target Normal Sheath Acceleration (TNSA) of protons is shown using a 1 $\times$ $10^{20}$ $W/cm^2$, 30 fs laser with intensity contrast better than $10^{-7}:1$. The optimum thickness was near 700 $nm$, resulting in a proton energy maximum of 24 $MeV$. Radiochromic film (RCF) was employed on both the laser and target normal axes, revealing minimal laser axis signal but a striking ring distribution in the low energy target normal ion signature that varies with liquid crystal thickness. Discussion of this phenomenon and a comparison to similar observations on other laser systems is included. 
\end{abstract}


\pacs{}

\maketitle 

\section{Introduction}

As the repetition rate of ultrashort (30 $fs$), ultra-intense ($I$ $>$ $10^{21}$ $W/cm^2$) laser systems increases to 10 $Hz$ and beyond,\cite{Gales15} several applications are being pursued with renewed interest. Among them is hadron cancer therapy,\cite{Bulanov04} which requires both higher energy ions and higher repetition rate target delivery than has currently been demonstrated. For this application the proton energy determines the depth of energy deposition of the ion beam, therefore understanding how to control the energy of laser-accelerated ions is critical. Investigations into the fundamental physics behind the various acceleration mechanisms promises other applications such as neutron beam generation, laboratory astrophysics, and positron production.\cite{Chen09, Roth09}

Currently studied ion acceleration mechanisms can be distinguished roughly by two experimental parameters: the target thickness and laser intensity contrast. Target Normal Sheath Acceleration (TNSA)\cite{Hatchett00, Snavely00} typically dominates for targets thicker than 1 $\mu m$ and for lasers with moderate to poor intensity contrast. The laser pre-pulses create a plasma at the front of the target from which electrons are accelerated by the main pulse; the resulting electric field at the target rear surface can accelerate ions at this location to tens of $MeV/nucleon$ energies in the target normal direction.

Higher contrast lasers and thinner targets can enable other mechanisms for relativistically intense lasers. Radiation Pressure Acceleration (RPA)\cite{Macchi10, Qiao12} and Break-Out Afterburner (BOA),\cite{Yin06} involve penetration of the laser past the classically expected plasma critical surface due to a relativistic modification to the electron plasma frequency. This relativistic transparency\cite{Palaniyappan12} can potentially accelerate the entire target volume along the laser axis, but only for sufficiently high contrast pulses. These newer acceleration methods have been studied in various simulations and experiments,\cite{Macchi13, Daido13} but details of their underlying physics and interplay with TNSA are still under investigation.

In addition to their energy spectrum, the spatial distribution of accelerated protons is also of interest in enabling applications and as it reveals physical processes governing the laser interaction and subsequent target evolution. For example, a ring of proton signal has been observed centered around the target normal direction for low energy ions, with several explanations as to the origin. It has been suggested that this feature is due to front surface generated shocks that accelerate ions from that location\cite{Haberberger12} or that penetrate to the rear target in time for the low energy proton acceleration,\cite{Xu06, Badziak08} or to an interaction between early accelerated protons and heavier ions accelerated via relativistic transparency effects pushing them from behind.\cite{Powell15, Dover16}

Discussed here are the energy spectra and spatial distributions obtained during a study comparing the target normal and laser axis accelerated ions from a variety of target thicknesses using thin films of liquid crystal. This material can be drawn into freely suspended films with thicknesses from 10 $nm$ to over 40 $\mu m$;\cite{Poole14} this range includes (and extends well above) that which is necessary for accessing any of the currently studied ion acceleration processes. The target formation technique will be presented along with experimental energy and spatial distribution data, and discussed in comparison to related results. 

\section{Liquid crystal film formation}

The formation process for freely suspending liquid crystal films utilizes the smectic mesophase of 4-octyl-4'-cyanobiphenyl (8CB), which intrinsically forms in stacked molecular layers, allowing thickness variation, and with sufficient surface tension to form planar films within rigid apertures.\cite{Poole14} Creating uniform thickness films and then controlling that thickness can be done by tuning film formation parameters such as temperature (near 28.5 $^{\circ}$C) and volume (on the order of 100 $nL$). 

For the experiment described here films were formed in 4 $mm$ diameter circular holes punched into copper plates roughly 5 $mm$ $\times$ 10 $mm$ $\times$ 1 $mm$ in dimension. These plate target frames were initially placed into a copper block heated by resistive heaters (25 $W$ maximum), and a type T thermocouple was used for temperature control and monitoring. Liquid crystal volume was applied to each frame individually with a precision syringe pump (Harvard Apparatus), allowed to equilibrate over several seconds, and then drawn across the aperture with a teflon-coated razor blade to minimize scratching to the copper surface. 

Film thickness was determined with a Filmetrics F20 multi-wavelength interference device, which measures thickness to within 2 $nm$. Films were formed repeatedly within a single frame until a desired thickness was achieved. If the correct volume and temperature parameters were used during the initial formation, the films could then be stored indefinitely: once the frame returned to room temperature the film would remain ``frozen'' at the formed thickness regardless of frame orientation, air currents, or even gross mechanical motion.

\section{Experimental setup}

\begin{figure} [ht]
	\includegraphics[width=0.5\textwidth]{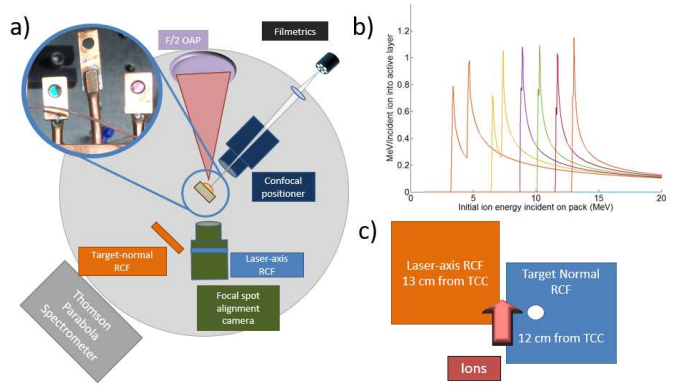}
	\caption{a) Schematic of experimental setup including focal spot alignment camera and confocal positioner for target alignment, Filmetrics device for liquid crystal film thickness measurement, and Thomson parabola spectrometer and radiochromic film (RCF) as ion diagnostics. b) Energy deposition into each RCF layer as a function of incident ion energy; MDV films have two separated active layers and hence two Bragg peaks, as in the two lowest energy curves here. c) Diagram of RCF spatial orientation in target chamber; the laser axis RCF was mounted on top of the focal spot camera and so was lowered into this position for a shot.}
	\label{fig:chambersetup}
\end{figure}

The experiment was performed on the Scarlet laser facility, which optimally yields 400 $TW$ from 12 $J$ in 30 $fs$, and is routinely capable of intensities above 5 $\times$ $10^{21}$ $W/cm^2$ due to a 2 $\mu m$ FWHM focal spot produced from an $F/2$ off-axis parabola. \cite{Scarlet16} For this experiment, reduced energy (5 $J$) and a defocused spot resulted in an intensity of $\sim$ 1 $\times$ $10^{20}$ $W/cm^2$. The experimental chamber setup is shown in Fig.\,\ref{fig:chambersetup}a, indicating the alignment and experimental diagnostics to be described below. The liquid crystal films were shot with $p$ polarization at an angle of incidence of 22.5$^{\circ}$. This was critical to determine the ion acceleration mechanism, since TNSA ions will travel along the target normal axis while mechanisms that require thinner targets are expected to produce ions that propagate along the laser axis.

Previous experiments on liquid crystals\cite{Poole14} revealed some difficulty in target alignment due to the uniform and transparent nature of the films. As such a target alignment procedure involving confocal microscopy was developed.\cite{Willis15} Here an alignment laser was tightly focused at normal incidence onto the film, and the reflection was imaged onto a 10 $\mu m$ single core fiber that served as a pinhole for the confocal system. Improper $z$ alignment would result in an increased spot size once the alignment beam was relayed back to the fiber such that the transmitted light signal was reduced. A variable gain photodiode was used to measure the reflected signal from thin, mostly transparent liquid crystal films as a function of target position. In this way the target $z$ position could be determined with $\pm 1 \mu m$ accuracy.

The primary experimental diagnostics were a compact-design Thomson parabola spectrometer\cite{Morrison11} placed along the target normal axis, and radiochromic film (RCF) stacks placed along both the target normal and laser axes. Here the target normal RCF had a 4 $mm$ hole punched to allow line of sight from the target to the Thomson parabola. RCF stacks utilized EBT, MDV2, and MDV3 films (Gafchromic/Ashland Specialty Ingredients) with appropriate thicknesses of copper and aluminum foil in between to allow measurement of protons from 1-30 $MeV$. The details of the RCF energy deposition curves and chamber placement are shown in Fig.\,\ref{fig:chambersetup}b and c. Energy deposition plots were calculated in the CSDA approximation in SRIM \cite{Ziegler10} using stoichiometric and thickness data from the film manufacturer.\cite{AshlandPC} When extracting data from films, the energy dose was corrected for under-response at the Bragg peak due to high linear energy transfer according to an empirical fit.\cite{Schollmeier14} The diagnostic utilized four or six RCF layers, spanning ion energy ranges up to 11 or 14 $MeV$, respectively. The laser axis RCF was mounted to a structure above the objective of the focal spot alignment camera such that lowering this to prevent damage during a shot brought the RCF into its appropriate position.

Liquid crystal films were formed in the copper frames (shown in inset of Fig. \ref{fig:chambersetup}a) on a setup bench to desired thicknesses. They were then transported to and installed within the experimental chamber target positioner. After chamber evacuation and target alignment the film thickness was verified with a Filmetrics device that had been set up outside a chamber port and optically relayed to the film. Not only did films typically survive installation, chamber evacuation, and target alignment, but also they maintained their original formation thickness.

\section{Results}

\begin{figure} [ht]
	\includegraphics[width=0.5\textwidth]{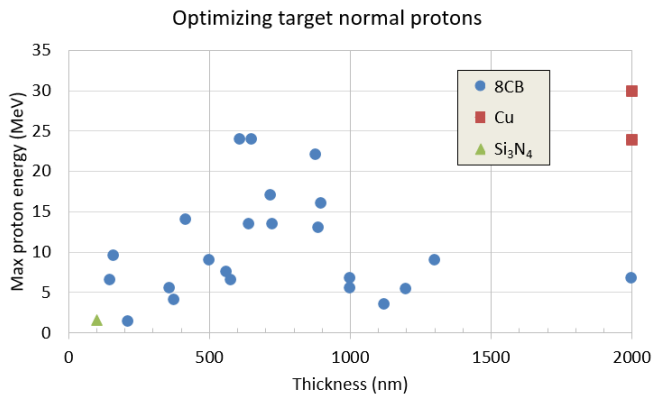}
	\caption{Graph of maximum proton energy recorded on the target-normal Thomson parabola spectrometer as a function of target thickness for 8CB liquid crystal, copper, and Si$_3$N$_4$ targets. An optimization of these TNSA protons is observed near 700 $nm$ for 8CB, resulting in 24 $MeV$ maximum energy protons.}
	\label{fig:thickscan}
\end{figure}

Figure\,\ref{fig:thickscan} shows the maximum proton energy along the target normal axis recorded on the Thomson parabola spectrometer as a function of target thickness. Most of these shots are from 8CB films, with a few 100 $nm$ Si$_3$N$_4$ and 2 $\mu m$ copper foil targets for comparison. A maximization of the TNSA proton energy can be seen around 700 $nm$ target thickness, with protons reaching 24 $MeV$--a factor of 2 optimization over the cutoff energy from thicknesses only 200 $nm$ different. Additionally, this proton energy from only 5 $J$ incident on target constitutes a high TNSA energy for a Ti:sapphire based system.\cite{Zeil10, Petrov16}

The decrease in proton energy for thicknesses below 700 $nm$ is attributed at least in part to insufficiently clean laser contrast. There was a pre-pulse 160 $ps$ prior to the main pulse seven orders of magnitude lower in intensity; the pre-plasma resulting from this would have affected thinner targets more strongly. The highest energy protons coming from 700 $nm$ thick targets is likely due to a trade-off between a thick enough target to survive the initial prepulse but also thin enough that electrons accelerated from the resulting pre-plasma have less time to diverge and therefore make a stronger sheath field at the target rear surface and also have more volume over their refluxing path where they are not within the target and can therefore contribute to surface sheath fields.\cite{Steinke10}

\begin{figure} [ht]
	\includegraphics[width=0.45\textwidth]{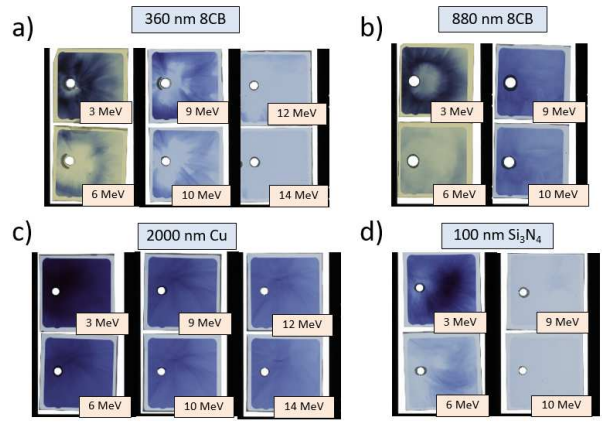}
	\caption{Target normal RCF stacks separated by Bragg peak energy for a variety of targets shot during this run, shown here to illustrate the ring structure observed in low energy protons. a) and b) show the feature in 360 $nm$ and 880 $nm$ 8CB targets, c) shows no ring structure in a 2 $\mu m$ Cu shot, and d) shows an accumulation of low energy dose on one side of a faint ring from a 100 $nm$ Si$_3$N$_4$ target.}
	\label{fig:RCFring}
\end{figure}

RCF data was also collected on target normal and laser axis. No appreciable laser axis signal was seen for any thickness, which is possibly attributable to the laser contrast inhibiting thin target acceleration. However, nearly every liquid crystal shot resulted in a ring of high density signal for the low energy protons, roughly centered on the target normal axis. An example RCF stack featuring this is shown in Fig.\,\ref{fig:RCFring}. This feature was not prominent in the 100 $nm$ Si$_3$N$_4$ nor seen at all in the 2 $\mu m$ Cu targets, nor were they affected by varying the pulse duration from 30 to 100 $fs$. 

Other experiments have shown this sort of annular feature in low energy protons onto RCF, with varying explanations. One possibility is that the ring of protons come from a different mechanism than TNSA, for example from front-surface collisionless shock acceleration.\cite{Haberberger12} However, this effect would be enhanced with a longer pulse, which would drive the front-side acceleration longer, but no such effect was observed when the pulse duration was increased here. Another possibility is that prepulse target heating causes an expansion of the rear surface,\cite{Badziak08} especially when it arrives $>$ 100 $ps$ prior to main pulse as was the case here. This shaped rear surface would give variation in target normal vector pointed radially outward from the laser spot location, and one which evolved over time as the target continued to expand. In this case, the largest rear surface deformation should occur at late times, which is when the slowest ions are accelerated, but one would expect the high energy ions to all be within the ring on later layers of the RCF since they are accelerated before this deformation has occurred. This was not the case for the high energy ions shown here, as will be shown below. Finally, recent work\cite{Powell15} suggests this ring is an artifact of relativistic transparency onset, as was shown on the 200 $J$ VULCAN laser for 40 $nm$ targets. While relativistic transparency is reasonable for those conditions, it is not expected for 700 $nm$ thick targets for the 5 $J$ energies used in this experiment.


To investigate the origin of this ring feature and its dependence on target parameters like thickness, a detailed RCF analysis was performed to combine the information recorded on the individual stacks into spatial maps of spectral information. This was done by binning the ions at energies corresponding to the peak response of each RCF layer.  Due to the broad high energy tail in the RCF response functions, the observed dose of each film can be considered as a sum of ions from its own bin and all higher energy bins, weighted by the film response function at each bin energy.  A weighted subtraction of the RCF film data then isolates the ion flux at each energy bin. 

For more closely spaced energy bins, recent unfolding techniques from Schollmeier \textit{et al.}\cite{Schollmeier14} provide more accurate ion spectra but require optimization, relying on minimization of higher order derivatives to avoid unphysical oscillation in the calculated spectra.  For the 4-6 energy bins in this experiment, the approximation used here avoids this oscillatory behavior, resulting in a more physically accurate spectral fit. As the bin energies are set at the response function maxima, results calculated with this method represent lower bounds.

\begin{figure} [ht]
	\includegraphics[width=0.45\textwidth]{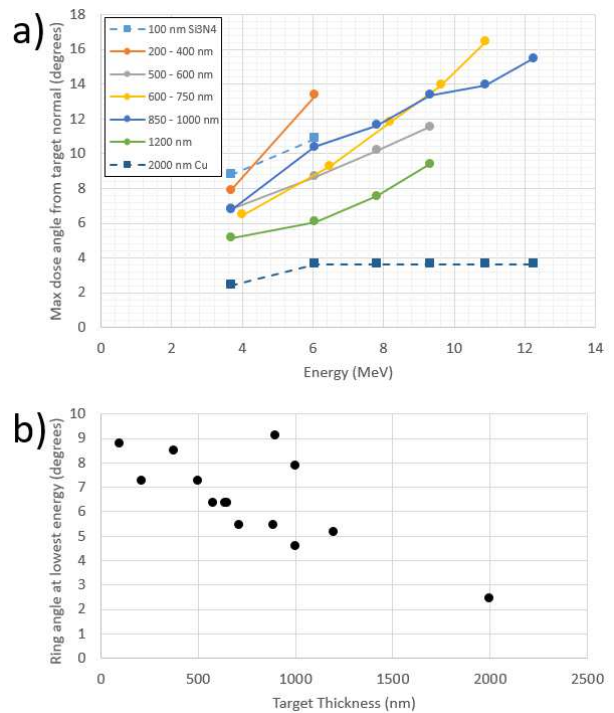}
	\caption{a) Angle of peak proton density as a function of energy (via the RCF layer position in the stack) for different thickness bins, with zero degrees being the target normal axis. The thinnest targets had no observable protons after the second RCF layer; for thicker liquid crystal targets (not copper) the ring feature grows with increasing energy at an equivalent rate for all thicknesses. Dashed lines represent non-liquid crystal targets. b) The initial angle of the proton ring feature (first non-saturated RCF layer) as a function of target thickness, showing a trend of larger initial angle for thinner targets.}
	\label{fig:anglethickness}
\end{figure}

A resulting spectral unfold plot is shown in Fig.\,\ref{fig:anglethickness}a, where the angle of the highest dose region on each RCF layer is recorded for a variety of thicknesses. Here 0$^{\circ}$ corresponds to the target normal axis as measured to be the center of the through hole punched to allow Thomson parabola spectrometer line of sight to target chamber center. All liquid crystal shots show the same slope of ring angle increasing at higher energies (deeper RCF layers). Figure\,\ref{fig:anglethickness}b plots the ring angle as observed on the first non-saturated RCF layer as a function of thickness, revealing that thicker targets generate rings with less divergence, possibly due to being less susceptible to prepulse-generated rear target expansion.\cite{Xu06}

\begin{figure} [ht]
	\includegraphics[width=0.45\textwidth]{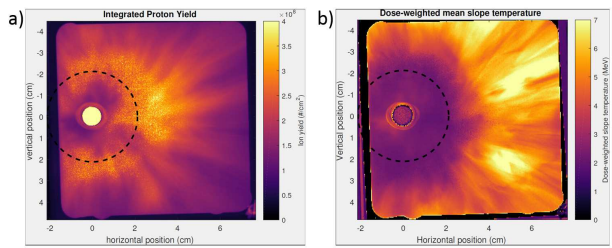}
	\caption{a) Integrated proton flux for a typical shot (890 $nm$) using the 2D spectral deconvolution as described in the text, showing the proton ring and high density proton signal on the target normal RCF pack. The dashed circle indicates 10$^{\circ}$ with respect to target normal. b) Processing of the position-dependent spectrum to yield a temperature-like parameter as a function of position. This shows the hottest ions falling outside the annular ring feature.}
	\label{fig:slopetemp}
\end{figure}

Additionally, the individual RCF layers were combined into an integrated ion flux spatial map as shown in Fig.\,\ref{fig:slopetemp}a. This was done by applying the spectral unfold described above as a function of position. The dose-weighted average ion slope temperature of Fig.\,\ref{fig:slopetemp}b was defined to be 

\begin{equation}
T_{slope} = \sqrt{\frac{-\mathcal{N}}{2\left\langle\frac{\partial N(\epsilon)}{\partial\epsilon}\right\rangle}_{\mathcal{N}}}, \quad \left\langle\frac{\partial N(\epsilon)}{\partial\epsilon}\right\rangle_{\mathcal{N}} = \frac{\int\frac{\partial N(\epsilon)}{\partial\epsilon}N(\epsilon)d\epsilon}{\int N(\epsilon)d\epsilon}
\end{equation}

where $N(\epsilon)$ is the extracted ion spectral function expressed in terms of ion energy $\epsilon$, and $\mathcal{N}$ is the total integrated ion flux at a given position. This expression allows an estimate of the slope temperature calculated using the highest flux ions, critically preventing lower-dose electron and x-ray background from having a dramatic effect on the calculated temperature.  This dose-weighted approximate slope temperature reduces to the real value in the case of a true Boltzmann spectrum input. The result shows that the highest temperature ions are outside the ring feature.

This analysis showing an increase of the ring angle with increasing ion energy is unexpected given previous explanations. A possible reconciliation is to assume that the lowest energy protons are those that can most easily be caught by heavier ions accelerated at later times of TNSA (expected because of their higher charge to mass ratio). The fastest carbon species, accelerated centrally from the target rear normal, could snowplow the slowest protons both forward and outward, coupling these two motions such that the most affected protons are both more energetic and diverging to a larger radius. The necessity of two ion species would explain why this result was seen for the liquid crystal targets but not for copper, which has a much lower population of carbon ions to affect protons (these coming only from the hydrocarbon contaminant layer on the back of the metal foil). This sort of multi-species interaction has been seen in some recent simulations,\cite{Padda16} with the key difference here being that relativistic effects were likely not present at this laser energy and contrast. Further simulations are required to ascertain the exact nature of the rear surface expansion and multi-species acceleration.

\section{Conclusion}

We have demonstrated the optimization of TNSA protons recorded along the target normal direction with both a Thomson parabola spectrometer and RCF, obtaining 24 $MeV$ with 5 $J$ of incident energy. Fine thickness variation was possible with liquid crystal films, and techniques adapted after this experiment now allow film formation in-situ at repetition rates surpassing once per minute.\cite{PooleAPL16} The proton spatial distribution has been analyzed from liquid crystal shots, showing possible evidence for the interplay between both target rear surface deformation and multi-species interaction during acceleration. Additional experiments and simulations are underway to investigate these findings.

\section*{Acknowledgments}

This work was supported by the DARPA PULSE program through a grant from AMRDEC and by the US Department of Energy under contract DE-NA0001976.

\bibliographystyle{unsrt}
\bibliography{LCionopt}

\end{document}